\documentclass[aps,graphicx,twocolumn]{revtex4}
\usepackage{graphicx,CJK}
\usepackage{subfigure}
\usepackage{color}
\usepackage{amsmath}
\usepackage{appendix}

\begin{document}


\title{Exact and efficient quantum simulation of open quantum dynamics \\ for various of Hamiltonians and spectral densities}

\author{Na-Na Zhang,$^{1,*}$ Ming-Jie Tao,$^{2,*}$ Wan-Ting He,$^{1,}$\footnote{These authors contributed equally to this work.} Xin-Yu Chen,$^{3}$\\
Xiang-Yu Kong,$^{3}$ Fu-Guo Deng,$^{1}$  Neill Lambert,$^{4}$ Qing Ai, $^{1,}$\footnote{aiqing@bnu.edu.cn}  Yuan-Chung Cheng$^{5,}$ \footnote{yuanchung@ntu.edu.tw}}

\address{$^{1}$Department of Physics, Applied Optics Beijing Area Major Laboratory,
Beijing Normal University, Beijing 100875, China\\
$^{2}$Space Engineering University, Beijing 101416, China\\
$^{3}$Department of Physics, Tsinghua University, Beijing 100084, China\\
$^{4}$Theoretical Quantum Physics Laboratory, RIKEN Cluster for Pioneering Research, Wako-shi, Saitama 351-0198, Japan\\
$^{5}$Department of Chemistry, National Taiwan University, Taipei City, Taiwan}

\date{\today}

\begin{abstract}
 Recently, we have theoretically proposed and experimentally demonstrated an exact and efficient quantum simulation of photosynthetic light harvesting in nuclear magnetic resonance (NMR), cf. B. X. Wang, \textit{et al.} npj Quantum Inf.~\textbf{4}, 52 (2018).
In this paper, we apply this approach to simulate the open quantum dynamics in various  photosynthetic systems with different Hamiltonians. By numerical simulations, we show that for Drude-Lorentz spectral density the dimerized geometries with strong couplings within the donor and acceptor clusters respectively exhibit significantly-improved efficiency. Based on the optimal geometry, we also demonstrate that the overall energy transfer can be further optimized when the energy gap between the donor and acceptor clusters matches the peak of the spectral density. Moreover, by exploring the quantum dynamics for different types of spectral densities, e.g. Ohmic, sub-Ohmic, and super-Ohmic spectral densities, we show that our approach can be generalized to effectively simulate open quantum dynamics for various Hamiltonians and spectral densities. Because $\log_{2}N$ qubits are required for quantum simulation of an $N$-dimensional quantum system, this quantum simulation approach can greatly reduce the computational complexity compared with popular numerically-exact methods.

\end{abstract}

\maketitle


\section{Introduction}
\label{sec:Introduction}

As commonly understood, the efficiency of exciton energy transfer (EET) in natural photosynthesis is close to unity \cite{Fleming94,Cheng09,Tao20}. Because of the discovery of EET with coherent features, the role of quantum coherence in EET efficiency has become a research hotspot in the past two decades \cite{Lambert13,Cao20,Engel07,Lee07,Wolynes09,Collini10,Hildner13,Tao16}.
Pigment-protein complexes in photosynthesis are essentially open quantum systems. Since the couplings between the system and the environment are of the same order of magnitude as the couplings within the system \cite{Cheng09,Tao20}, non-Markovian features arise and make simulating the open quantum dynamics therein difficult \cite{Breuer07,Breuer16,Vega17,Li18,Ishizaki09-1}. So far, some theoretical methods have been proposed to effectively simulate EET in photosynthesis \cite{Tao20}, such as the numerically-exact hierarchical equation of motion (HEOM) \cite{Tanimura06,Ishizaki09-2,Yan04,Zhou05,Shao04,Tang15,Liu14,Schroder07}, the quantum jump approach \cite{Olaya-Castro08,Ai14}, the small-polaron quantum master equation \cite{Jang08}, the modified Redfield theory and its coherent generalization \cite{Yang02,Hwang-Fu15} and so on. Among these methods simulating the EET, the HEOM yields exact quantum dynamics in the whole parameter regime, e.g. the F\"{o}rster regime and the Redfield regime \cite{Ishizaki09-2,Tao20}. It is helpful to revealing the role of quantum coherence in optimizing the photosynthetic EET \cite{Lambert13} and clearly elucidating the design principals of artificial light-harvesting devices \cite{Dong12,Mostarda13,Knee17,Zech14,Xu18}.

However, despite the fact the that  HEOM has been widely used in the study of open quantum dynamics, including EET in natural photosynthesis, in the case of large dimensions and complex spectral densities, the numerical overhead becomes very large.
Recently, we proposed a novel experimental approach to \textit{exactly and efficiently} simulate EET in photosynthesis~\cite{Wang18}. We generate a large number of realizations driven by random Hamiltonians, and by averaging over the ensemble we obtain a density matrix whose dynamics is subject to decoherence. As a demonstration, we adopted a prototype in Ref.~\cite{Ai13} and compared the results of nuclear magnetic resonance (NMR) simulation and HEOM calculation under Drude-Lorentz noise. We showed that it is valid to efficiently simulate the exact quantum dynamics in the photosynthetic EET by using NMR if the number of random realizations is sufficiently large.

As we know, for systems with large dimension or complex spectral densities, the HEOM requires a huge amount of computation resources. For example, to simulate an $N$-level system, the computation cost of an $\mathcal{N}$-layer HEOM scales exponentially in $\mathcal{N}$ ($\mathcal{N}\leq N$) \cite{Shi09}. However, in the quantum simulation, because the quantum dynamics of $N$ states can be simulated by using $\log_{2}N$ qubits, the computation cost is a polynomial of $N$.
Therefore, this quantum simulation can effectively reduce the computational complexity.

In 2013, it was demonstrated that the efficiency of energy transfer can be improved when there is strong coupling within donor and acceptor pairs by studying energy transfer in a linear-tetramer model \cite{Ai13}. In the same year, del Rey \textit{et al.} proposed a design principle called phonon antenna. By spectrally sampling optimum in their local environmental fluctuations, the coherence between internal pigments can affect and optimize the way excitation flows \cite{Rey13}. And the strong coupling to an under-damped vibrational mode can help the photosynthetic complex to overcome the energy barrier between the donor and acceptor, and thus increase the efficiency \cite{Gorman18}.

\color{black}
In Ref.~\cite{Wang18}, only a quantum simulation with a specific Hamiltonian and Drude-Lorentz noise was demonstrated. The clustered geometry was proven to be optimal for a broad parameter regime by the coherent modified Redfield theory \cite{Ai13}. However, the theory is shown to break down when the reorganization energy is much larger than the intra-system coupling \cite{Hwang-Fu15,Chang15}. A natural question arises: does the above discovery still hold in a broad range of parameters by a numerically-exact approach? On the other hand, since both modifying the geometry and spectral-sampling in local environmental fluctuations can optimize energy flow in EET respectively, it is quite natural to ask whether light harvesting can be further optimized if both means have been applied?


The structure of this paper is organized as follows.
In Sec.~\ref{sec:NMR}, we give detailed introduction to this approach for quantum simulation in NMR, which utilizes the bath-engineering technique \cite{Soare14-1,Soare14-2} and the gradient ascent pulse engineering (GRAPE) algorithm \cite{Khaneja05,Li17}. In Sec.~\ref{sec:NCA}, we first consider the simulation of the EET dynamics in a linear-tetramer model with various of Hamiltonians. 
After the Hamiltonian is optimized, we also consider how to improve the EET efficiency from the aspect of the spectral density. Through this specific model, we confirm the discovery in Ref.~\cite{Rey13}. 
 Besides, in Sec.~\ref{sec:NCA} we also consider the effects of different types of spectral densities on the EET dynamics, which was not analyzed in Ref.~\cite{Wang18}.
In Sec.~\ref{sec:NCA}, we compare the computational complexities of NMR simulation and HEOM, and analyze the possible errors in NMR simulations.
Finally, we discuss the prospect and conclusions in Sec.~\ref{sec:Conclusion}.

\section{Theory for Quantum Simulation}
\label{sec:NMR}

In this section, we introduce how to simulate the EET dynamics of photosynthesis in NMR systems. In Ref.~\cite{Wang18}, we simulated the influence of noise by using the bath-engineering technique \cite{Soare14-1,Soare14-2}, and realized the evolution of quantum states by using the GRAPE algorithm \cite{Khaneja05,Li17}. In the following, before we summarize these two techniques, we shall give a brief introduction to our model system and the HEOM theory for photosynthetic EET.

\subsection{Model Photosynthetic System}

\begin{figure}[h]
\includegraphics[width=7.6cm]{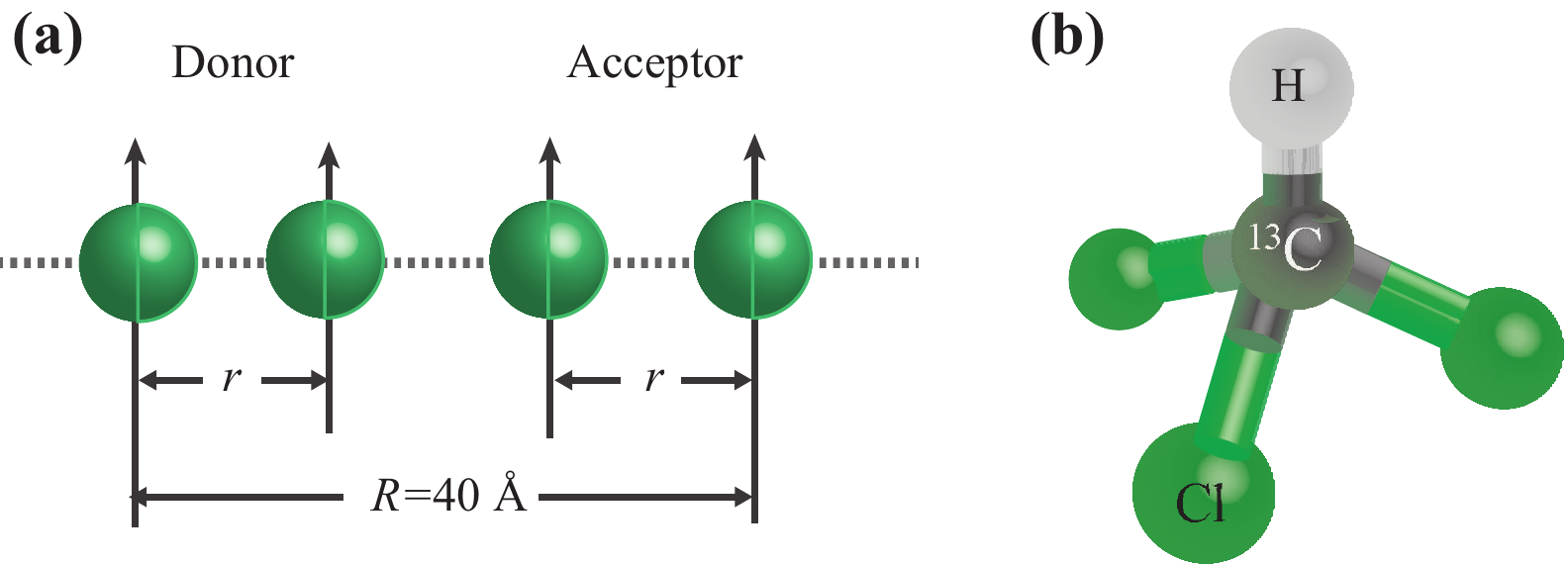}
\caption{Photosynthetic tetramer and physical system for NMR simulation.
(a) Linear geometry with four chromophores for photosynthetic EET. The distance between the first donor and the last acceptor is fixed as $R=40~\rm{\mathring A}$. We assume that the distances within the pairs of donors and acceptors, i.e. $r$, are equal. All of the transition dipoles are perpendicular to the horizontal axis.
(b) Chemical structure of a $^{13}\textrm{C}$-labeled chloroform molecule, where H and $^{13}\textrm{C}$ nuclear spins are chosen as the two qubits for quantum simulation.}\label{fig1}
\end{figure}

Following Ref.~\cite{Wang18}, we use the tetramer model, which contains four chlorophyll molecules, for the quantum simulation. As shown in Fig.~\ref{fig1}(a),
the left pair of chlorophylls act as donors and the right pair as acceptors.
We adopt the Frenkel-exciton Hamiltonian \cite{Cheng09,Valkunas13}
\begin{equation}
H_{\rm{EET}}=\sum_{i=1}^{4}\varepsilon_{i}\vert i\rangle\langle i\vert +\sum_{i\neq j=1}^{4}J_{ij}\vert i\rangle\langle j\vert
\end{equation}
to describe the dynamics of EET in photosynthesis. Here $\vert i\rangle~(i=1,2,3,4)$ represents that only $i$th molecule is in the excited state while the others are in the ground state. $\varepsilon_{i}$ is the site energy of $i$th exciton.
For simplicity, the electronic interaction between $i$th and $j$th excitons is given by the dipole-dipole interaction as
\begin{equation}
J_{ij}=\frac{1}{4\pi\varepsilon_{0}r_{ij}^{3}}\left[\vec{\mu}_{i}\cdot\vec{\mu}_{j}
-3(\vec{\mu}_{i}\cdot\hat{r}_{ij})(\vec{\mu}_{j}\cdot\hat{r}_{ij})\right],\label{eq:J}
\end{equation}
where $\vec{r}_{ij}=r_{ij}\hat{r}_{ij}$ is the displacement vector from site $i$ to site $j$, $\vec{\mu}_{i}$ the transition dipole of site $i$, $\varepsilon_{0}$ the vacuum permittivity. In numerical simulations, we take $\mu_{j}=7.75~\textrm{D}$  and $r\in[6,14]~\rm{\mathring A}$, which are typical in natural photosynthesis.

For photosynthetic complexes with $N$ chlorophylls, there are only $N$ single-excitation states in the process of energy transfer. As a result, only $\log_{2} N$ qubits are required for quantum simulation and thus two qubits are needed to simulate the tetramer model.
In this case, the single-excitation states in photosynthesis are encoded as two-qubit product states as $\vert 1 \rangle=\vert 00 \rangle$, $\vert 2 \rangle=\vert 01 \rangle$, $\vert 3 \rangle=\vert 10\rangle$, and $\vert 4 \rangle=\vert 11 \rangle$.
As shown in Fig.~\ref{fig1}(b), we regard H and C nuclear spins as the two qubits.
The Hamiltonian $H_{\rm{NMR}}$ implemented in NMR simulations is $H_{\rm{NMR}}=H_{\rm{EET}}/C$ with a scaling factor
\begin{equation}
C=3\times10^{9}.
\end{equation}

In photosynthetic complexes, the energy transfer is assisted by the interaction between the system and the bath, which can be described as
\begin{equation}
H_{\rm{SB}}=\sum_{i,k}g_{ik}V_j\left(a_{ik}^{\dagger}+a_{ik}\right).\label{eq:Hsb}
\end{equation}
Here, $V_j=\vert i\rangle\langle i\vert$, $a_{ik}^{\dagger}$ ($a_{ik}$) is the creation (annihilation) operator of the $k$th phonon mode of the $i$th molecule, and $g_{ik}$ represents the coupling strength.
$H_{\rm{SB}}$ is the main cause of energy relaxation in photosynthetic systems.
All the information about the couplings between system and environment can be given by the spectral density, i.e.,
\begin{equation}
G_{\rm{EET}}(\omega)=\sum_{k}g_{ik}^{2}\delta(\omega-\omega_{k}).
\end{equation}
The spectral density plays an important role in the optimization of EET in photosynthesis. It has been shown that the energy transfer can be improved by adjusting the parameters of photosynthetic system so that the energy gap matches the optimal frequency of the spectral density \cite{Rey13}.

\subsection{Hierarchical Equation of Motion Method}

The exact EET dynamics can be given by the HEOM \cite{Ishizaki09-2,Ishizaki09-3}
\begin{equation}
\frac{\partial}{\partial t} \sigma_{\vec{n}}\!\!=\!\!(-i\mathcal{L}_e+\sum_j n_j\gamma_j)\sigma_{\vec{n}}
+\sum_j\Phi_j\sigma_{\vec{n}_{j+}}+\sum_j n_j\Theta_j\sigma_{\vec{n}_{j-}},
\end{equation}
where $\sigma_{\vec{n}}$ and $\sigma_{\vec{n}_{j\pm}}$
are the auxiliary density matrices with $\vec{n}=(n_1,n_2,\cdots,n_j,\cdots)$ and $\vec{n}_{j\pm}=(n_1,n_2,\cdots,n_j\pm1,\cdots)$,
 and $n_j$'s non-negative integers, $\sigma_{\vec{0}}=\rho$ the reduced density matrix of photosynthetic system, $\mathcal{L}_e$ the Liouville superoperator of $H_\textrm{EET}$.
Besides, $\Phi_j=iV_j^{\times}$ and $\Theta_j=i(\frac{2\lambda_j}{\beta\hbar^2}V_j^{\times}-i\frac{\lambda_j}{\hbar}V_j^{\circ})$
with $V_j^{\times}\sigma_{\vec{n}}=V_j\sigma_{\vec{n}}-\sigma_{\vec{n}}V_j$ and $V_j^{\circ}\sigma_{\vec{n}}=V_j\sigma_{\vec{n}}+\sigma_{\vec{n}}V_j$,
where $V_j=\vert j\rangle\langle j\vert$.
\color{black}
For Drude-Lorentz spectral density, i.e.,
\begin{equation}
G_\textrm{EET}(\omega)=\frac{2\lambda_j\gamma_j\omega}{\omega^2+\gamma_j^2}\label{eq:Geet}
\end{equation}
with $\lambda_j$ the reorganization energy and $\gamma_j$ the relaxation rate.
For a generic spectral density, it can be decomposed into a summation of Lorentzian form \cite{Meier99,Jiang18}.

\subsection{The process of quantum simulation}

In the quantum simulation of photosynthetic EET with NMR system, there are three steps: First of all, we initialize the system for quantum simulation, i.e., preparation of pseudo-pure states. Afterwards, the system evolves under the evolution operator $U$. Finally, the state of the system is measured. In the following, we shall discuss the details of the three steps.

\subsubsection{Initialization of Quantum States}

At the room temperature, the two-qubit system is initially in thermal-equilibrium state, i.e., \cite{Chuang98-1}
\begin{equation}
\rho_{\rm{eq}}\approx\frac{1}{4}\sigma^{(0)}_1\sigma^{(0)}_2+\epsilon(\gamma_{\rm{H}}\sigma^{(3)}_1\sigma^{(0)}_2+\gamma_{\rm{C}}\sigma^{(0)}_1\sigma^{(3)}_2),
\end{equation}
where $\sigma^{(0)}_j$ and $\sigma^{(\alpha)}_j$ ($\alpha=1,2,3$) are respectively the unit matrix and Pauli operators of qubit $j$, $\epsilon\approx1.496\times10^{-13}~\textrm{rad}^{-1}\cdot\textrm{s}\cdot\textrm{T}$ characterizes polarization,  $\gamma_{\rm{H}}=2.675\times10^8~\textrm{rad}\cdot\textrm{s}^{-1}\cdot\textrm{T}^{-1}$ and $\gamma_{\rm{C}}=6.726\times10^7~\textrm{rad}\cdot\textrm{s}^{-1}\cdot\textrm{T}^{-1}$ correspond respectively to the gyromagnetic ratios of the ${}^{1}\rm{H}$ and ${}^{13}\rm{C}$ nuclei \cite{Vandersypen04}.
In the experiment, by using the spatial average method \cite{Chuang98-2,Cory98}, the system is prepared in the pseudo-pure state as
\begin{equation}
\rho_{00}=\frac{1-\delta}{4}\sigma^{(0)}_1\sigma^{(0)}_2+\delta\vert 00\rangle\langle 00\vert,
\end{equation}
where $\delta\simeq10^{-5}$.
Since the unitary evolutions and measurements have no effects on the unit matrix part, the final results of the experiments are only influenced by the second part, i.e., $\vert 00\rangle$.

\subsubsection{Evolution under Hamiltonian with Noise}

The total Hamiltonian for simulating EET process is
\begin{eqnarray}
H(t)&=&H_{\rm{NMR}}+H_{\rm{PDN}}.
\end{eqnarray}
$H_{\rm{PDN}}$ is introduced to mimic the effects of the local baths in photosynthesis. By the bath-engineering technique \cite{Soare14-1,Soare14-2,Zhen16}, various types of spectral densities can be effectively simulated. The system Hamiltonian $H_{\rm{NMR}}$ can be obtained by the photosynthetic Hamiltonian scaled by the factor $C$.
In the quantum simulation, the evolution is divided into $L$ steps with the total evolution time $t=L\Delta t$, and the evolution operator is
\begin{equation}
U(t)=\prod_{j=1}^{L}U_{j}=\prod_{j=1}^{L}\exp[-iH_{j}\Delta t],
\end{equation}
where $H_{j}=H(t_{j})$ is the Hamiltonian at $t_{j}=j\Delta t$. Here, $U(t)$ can be decomposed into a series of experimentally-feasible pulses by the GRAPE algorithm \cite{Khaneja05,Li17}. In the following, we will introduce the bath-engineering technique and the GRAPE algorithm in detail.\\

\begin{center}
\textit{\small{2.1   Bath Engineering}}\\
\end{center}

In the process of EET, the system generally interacts with its environments.
In the NMR systems, by artificially injecting noise, the impact of the environment is effectively simulated. This bath-engineering technique has been successfully realized in ion traps and NMR \cite{Soare14-1,Soare14-2,Zhen16,Wang18}.

In order to mimic the system-bath interaction in Eq.~(\ref{eq:Hsb}), we utilize a dephasing noise which comes from the inhomogeneous and non-static magnetic fields in the NMR systems. The Hamiltonian of the dephasing noise is
\begin{equation}
H_{\rm{PDN}}=\sum_{m=1,2}\vec{B}_m(t)\cdot\vec{\sigma}_m,
\end{equation}
which relies on generating stochastic errors by performing phase modulations on a constant-amplitude carrier, i.e.,
\begin{eqnarray}
\vec{B}_m(t)&=&\Omega_{m}\cos\left[\omega_{m}t+\phi_{m}(t)\right]\hat{z},   \\
\phi_{m}(t)&=&\sum_{j=1}^{N_c}\alpha^{(m)}_{z}F(\omega_{j})\sin(\omega_{j}t +\phi^{(m)}_{j}),
\end{eqnarray}
where $\Omega_{m}$ is the constant amplitude of a magnetic field with driving frequency $\omega_{m}$, $\phi_{m}(t)$ the time-dependent phase with
$\alpha^{(m)}_{z}$ the amplitude of noise and $\phi^{(m)}_{j}$ a random number,
$\omega_{j}=j\omega_{0}$ with $N_c\omega_{0}$ and $\omega_{0}$ being cutoff and base frequencies, respectively.

The derivative of $\phi_{m}$ with respect to the time is
\begin{equation}
\beta_{m}(t)=\frac{d\phi_{m}(t)}{dt}=\sum_{j=1}^{N}\alpha^{(m)}_{z}F(\omega_{j})\omega_{j}
\cos(\omega_{j}t +\phi^{(m)}_{j}),
\end{equation}
which describes the distribution of noise in the time domain.
The second-order correlation function of $\beta_{m}(t)$ is
\begin{eqnarray}
\langle\beta_{m}(t+\tau)\beta_{m}(t)\rangle\!\!\!&=&\!\!\!\lim_{T\to\infty}\frac{1}{2T}
\int_{-T}^{T}dt\beta_{m}(t+\tau)\beta_{m}(t) \label{eq:CorreFunc}\\
\!\!\!&=&\!\!\!(\frac{\alpha^{(m)}_{z}}{2})^{2}\sum_{j}[\omega_{j}F(\omega_{j})]^{2}
(e^{i\omega_{j}\tau}+e^{-i\omega_{j}\tau}).\nonumber
\end{eqnarray}
By Fourier transform of the above equation, the power spectral density of the noise can be obtained as
\begin{eqnarray}
S_{m}(\omega)\!\!\!&=&\!\!\!\int_{-\infty}^{\infty}d\tau e^{-i\omega\tau}\langle\beta_{m}(t+\tau)\beta_{m}(t)\rangle\nonumber  \\
\!\!\!&=&\!\!\!\frac{\pi(\alpha^{(m)}_{z})^{2}}{2}\sum_{j=1}^{N}[\omega_{j}F(\omega_{j})]^{2}
[\delta(\omega-\omega_{j})\!\!+\!\!\delta(\omega+\omega_{j})].\nonumber  \\
\end{eqnarray}
In obtaining Eq.~(\ref{eq:CorreFunc}), we assume the average over the ensemble is equivalent to the average over the time, which is valid in the large-ensemble limit \cite{Goodman15}.

The noise is essentially a stochastic process and can be characterized by its power spectral density.
In general, we first map the spectral density of photosynthesis to the power spectral density of noise. Then, we inversely derive the modulation function $F(\omega_{j})$. and thus the time-domain function of noise $\beta_m(t)$. Obviously, the type of the noise is determined by the power spectral density and thus the function $F(\omega_{j})$.\\

\begin{center}
\textit{\small{2.2  Gradient Ascent Pulse Engineering Algorithm}}\\
\end{center}

The GRAPE algorithm has become the most commonly-used optimal-control theory for unitary evolutions in NMR systems \cite{Khaneja05,Li17}.
For an $N$-qubit NMR system, the total Hamiltonian $H_{\rm{tot}}$ includes the internal term $H_{\rm{int}}$ and the radio-frequency (RF) term $H_{\rm{RF}}$, and thus reads
\begin{eqnarray}
H_{\rm{tot}}&=&H_{\rm{int}}+H_{\rm{RF}}, \\
H_{\rm{RF}}&=&-\sum_{k=1}^{2}\gamma_{k}B_{k}[\cos(\omega^{\rm{RF}}_{k}t+\phi^{\rm{RF}}_{k})\sigma^{(1)}_{k}\nonumber\\
&&+\sin(\omega^{\rm{RF}}_{k}t+\phi^{\rm{RF}}_{k}\sigma^{(2)}_{k})],
\end{eqnarray}
where $B_{k}$, $\omega^{\rm{RF}}_{k}$ and $\phi^{\rm{RF}}_{k}$ are the amplitude, driving frequency and phase of the control field on the $k$th nuclear spin with gyromagnetic ratio $\gamma_{k}$, respectively.

The purpose of the GRAPE algorithm is to design a unitary evolution $U_{D}$ by iteration to make it very close to the target evolution $U_{T}$, so as to find the optimal amplitudes $B_{k}$ and phases $\phi^{\rm{RF}}_{k}$ of the control fields.
The fidelity of $U_{D}$ relative to $U_{T}$ can be expressed as
$F=\vert\textrm{Tr}(U_{T}^{\dagger}U_{D})\vert/2^{2}$.
We assume that the total evolution time is $T$ and is divided into $N$ steps, i.e., $\Delta t=T/N$. And the amplitudes and phases of the control fields within each step are constant.
Thus, in $j$th step, the time evolution operator of the system can be expressed as
\begin{equation}
U_{j}=e^{-i\Delta t\left[H_{\rm{int}}+\sum_{k=1}^{2}\sum_{\alpha=1}^{2}u^{(\alpha)}_{k}(j)\sigma^{(\alpha)}_{k}
\right]},
\end{equation}
where $u^{(1)}_{k}(j)=\gamma_{k}B_{k}\cos(\omega^{\rm{RF}}_{k}t_j+\phi^{\rm{RF}}_{k})$ and $u^{(2)}_{k}(j)=\gamma_{k}B_{k}\sin(\omega^{\rm{RF}}_{k}t_j+\phi^{\rm{RF}}_{k})$ are assumed to be constant.
The total evolution operator of the system is $U_{D}=U_{N}U_{N-1}\dots U_{2}U_{1}$.
By calculating the derivative of the fidelity $F$ with respect to $u^{(\alpha)}_{k}(j)$, we can obtain
\begin{align}
g^{(\alpha)}_{k}(j)&=\frac{\partial F}{\partial u^{(\alpha)}_{k}(j)}\nonumber \\
&\approx -\frac{2}{2^{n}}\textrm{Re}[U_{T}^{\dagger}U_{N}\dots U_{j+1}(-i\Delta t\sigma^{(\alpha)}_{k})U_{j}\dots U_{1}].
\end{align}
Afterwards, we replace $u^{(\alpha)}_{k}(j)$ by $u^{(\alpha)}_{k}(j)+\epsilon_s g^{(\alpha)}_{k}(j)$ with $\epsilon_s$ the iteration step. By repeating the above steps, we will find that the fidelity is increasing gradually.
To summarize the general steps of the GRAPE algorithm, we set an initial value of $u^{(\alpha)}_{k}(j)$, and calculate the derivative of the fidelity $g^{(\alpha)}_{k}(j)$, and iterate until the fidelity changes less than the selected threshold. After terminating the algorithm, we shall perform the measurements to obtain the final result.

\subsubsection{Tomography}
\label{subsec:Tomography}

In NMR, the free-induction decay (FID) signal is employed to measure the density matrix of the final state \cite{Vandersypen04,Lee02,Lu16,Xin17}, i.e.,
\begin{equation}
S^{U}(t)\propto\textrm{Tr}[e^{-iH_\textrm{int}t}U\rho U^{\dagger} e^{iH_\textrm{int}t}\sum_{k=1}^2(\sigma_k^{(1)}-i\sigma_k^{(2)})],
\end{equation}
where $\rho$ is the density matrix after the above quantum simulation approach has been applied,
\begin{equation}
H_\textrm{int}=\frac{\omega_1}{2}\sigma_1^{(3)}+\frac{\omega_2}{2}\sigma_2^{(3)}+\frac{\pi J}{2}\sigma_1^{(3)}\sigma_2^{(3)}
\end{equation}
is the internal Hamiltonian with $\omega_1=\gamma_\textrm{C}B$, $\omega_2=\gamma_\textrm{H}B$, and $J=215.1$~Hz \cite{Chuang98-1}.
All the elements of the density matrix can be given in terms of the expectations of the 16 observables $\{\sigma^{(i)}_1\otimes\sigma^{(j)}_2\}~(i,j=0,1,2,3)$.
In this paper, we mainly focus on the diagonal elements of the density matrix, i.e., \begin{align}\label{EQ:pho}
\rho_{11}&=\frac{1}{4}[1+\langle \sigma^{(3)}_1\sigma^{(0)}_2\rangle+\langle \sigma^{(0)}_1\sigma^{(3)}_2\rangle+\langle \sigma^{(3)}_1\sigma^{(3)}_2\rangle], \notag\\
\rho_{22}&=\frac{1}{4}[1+\langle \sigma^{(3)}_1\sigma^{(0)}_2\rangle-\langle \sigma^{(0)}_1\sigma^{(3)}_2\rangle-\langle \sigma^{(3)}_1\sigma^{(3)}_2\rangle], \\
\rho_{33}&=\frac{1}{4}[1-\langle \sigma^{(3)}_1\sigma^{(0)}_2\rangle+\langle \sigma^{(0)}_1\sigma^{(3)}_2\rangle-\langle \sigma^{(3)}_1\sigma^{(3)}_2\rangle],\notag\\
\rho_{44}&=\frac{1}{4}[1-\langle \sigma^{(3)}_1\sigma^{(3)}_2\rangle-\langle \sigma^{(0)}_1\sigma^{(3)}_2\rangle-\langle \sigma^{(3)}_1\sigma^{(0)}_2\rangle)].\notag\label{eq:diagRho}
\end{align}
We also provide the expressions for the other elements of the density matrix in Eq.~(\ref{eq:A1}) of Appendix~\ref{sec:AppendixA}.

In order to obtain the diagonal elements of the density matrix, we only need to apply a unitary operation $U$ to the system, which is $U=\{
\sigma_1^{(0)} \otimes e^{-i\frac{\pi}{4}\sigma_2^{(2)}},
e^{-i\frac{\pi}{4}\sigma_1^{(2)}} \otimes \sigma_2^{(0)}\}$.
For example, when $U=e^{-i\frac{\pi}{4}\sigma_1^{(2)}} \otimes \sigma_2^{(0)}$ is applied on the system before the measurement,
the FID signal reads
\begin{equation}
S^{\textrm{YI}}(t)\propto \sum_{k=1}^{2}\sum_{p=0}^{1}S_{kp}^{\textrm{YI}}e^{i[\omega_k+(-1)^p\pi J]t},\label{eq:Syi}
\end{equation}
where in the superscript $I=\sigma^{(0)}$ and $Y=\sigma^{(2)}$,
\begin{eqnarray}
S_{10}^{\textrm{YI}}\!\!=\!\!\langle\sigma_1^{(3)}\sigma_2^{(0)}\rangle+\langle\sigma_1^{(3)}\sigma_2^{(3)}\rangle
+i(\langle\sigma_1^{(2)}\sigma_2^{(0)}\rangle+\langle\sigma_1^{(2)}\sigma_2^{(3)}\rangle),\notag\\
S_{11}^{\textrm{YI}}\!\!=\!\!\langle\sigma_1^{(3)}\sigma_2^{(0)}\rangle-\langle\sigma_1^{(3)}\sigma_2^{(3)}\rangle
+i(\langle\sigma_1^{(2)}\sigma_2^{(0)}\rangle-\langle\sigma_1^{(2)}\sigma_2^{(3)}\rangle),\notag\\
S_{20}^{\textrm{YI}}\!\!=\!\!\langle\sigma_1^{(0)}\sigma_2^{(1)}\rangle-\langle\sigma_1^{(1)}\sigma_2^{(1)}\rangle
+i(\langle\sigma_1^{(0)}\sigma_2^{(2)}\rangle-\langle\sigma_1^{(1)}\sigma_2^{(2)}\rangle),\notag\\
S_{21}^{\textrm{YI}}\!\!=\!\!\langle\sigma_1^{(0)}\sigma_2^{(1)}\rangle-\langle\sigma_1^{(1)}\sigma_2^{(1)}\rangle
+i(\langle\sigma_1^{(0)}\sigma_2^{(2)}\rangle-\langle\sigma_1^{(1)}\sigma_2^{(2)}\rangle).\notag\\
\label{eq:diagObser}
\end{eqnarray}
By Fourier transform of Eq.~(\ref{eq:Syi}), we can obtain the above four quantities.
The expectations of the observables contained in Eq.~(\ref{EQ:pho}) can be written in terms of the FID signals as
\begin{subequations}
\begin{align}
\langle\sigma^{(0)}_1\sigma^{(3)}_2\rangle=\frac{\eta}{2}[\textrm{Re}(S_{20}^{\textrm{IY}})+\textrm{Re}(S_{21}^{\textrm{IY}})],\\
\langle\sigma^{(3)}_1\sigma^{(0)}_2\rangle=\frac{\eta}{2}[\textrm{Re}(S_{10}^{\textrm{YI}})+\textrm{Re}(S_{11}^{\textrm{YI}})],\\
\langle\sigma^{(3)}_1\sigma^{(3)}_2\rangle=\frac{\eta}{2}[\textrm{Re}(S_{20}^{\textrm{IY}})-\textrm{Re}(S_{21}^{\textrm{IY}})].
\end{align}
\end{subequations}
Here, $\textrm{Re}(x)$ and $\textrm{Im}(x)$ are respectively the real and imaginary parts of $x$, and $\eta$ is a constant which replies on experimental details such as a
receiver gain and the amount of spins \cite{Lee02}.
The method for measuring the other elements of the density matrix is also supplemented in
Appendix~\ref{sec:AppendixA}.

%

\section{NUMERICAL CALCULATION AND ANALYSIS }
\label{sec:NCA}

In Ref.~\cite{Wang18}, only the EET dynamics for a specific Hamiltonian under the Drude-Lorentz noise was demonstrated. In this section, we will use the above quantum simulation approach to investigate the dynamics for different Hamiltonians and various types of spectral densities. In addition, we will compare the computational complexity of the quantum simulation and the HEOM, and analyze the errors in the quantum simulation.

\subsection{Hamiltonians}

In Ref.~\cite{Ai13}, the effects of the geometry on the energy transfer was investigated. In the study, it was revealed that the dimerized structure explores coherent relaxation to promote the energy transfer within the dimer. Since the coupling between two sites is significantly dependent on the distance between two chlorophylls, different geometries correspond to different Hamiltonians. In this sub-section, we analyze the EET dynamics for different Hamiltonians with the quantum simulation approach.

\begin{figure*}[htbp]
\centering
\includegraphics[scale=1.1]{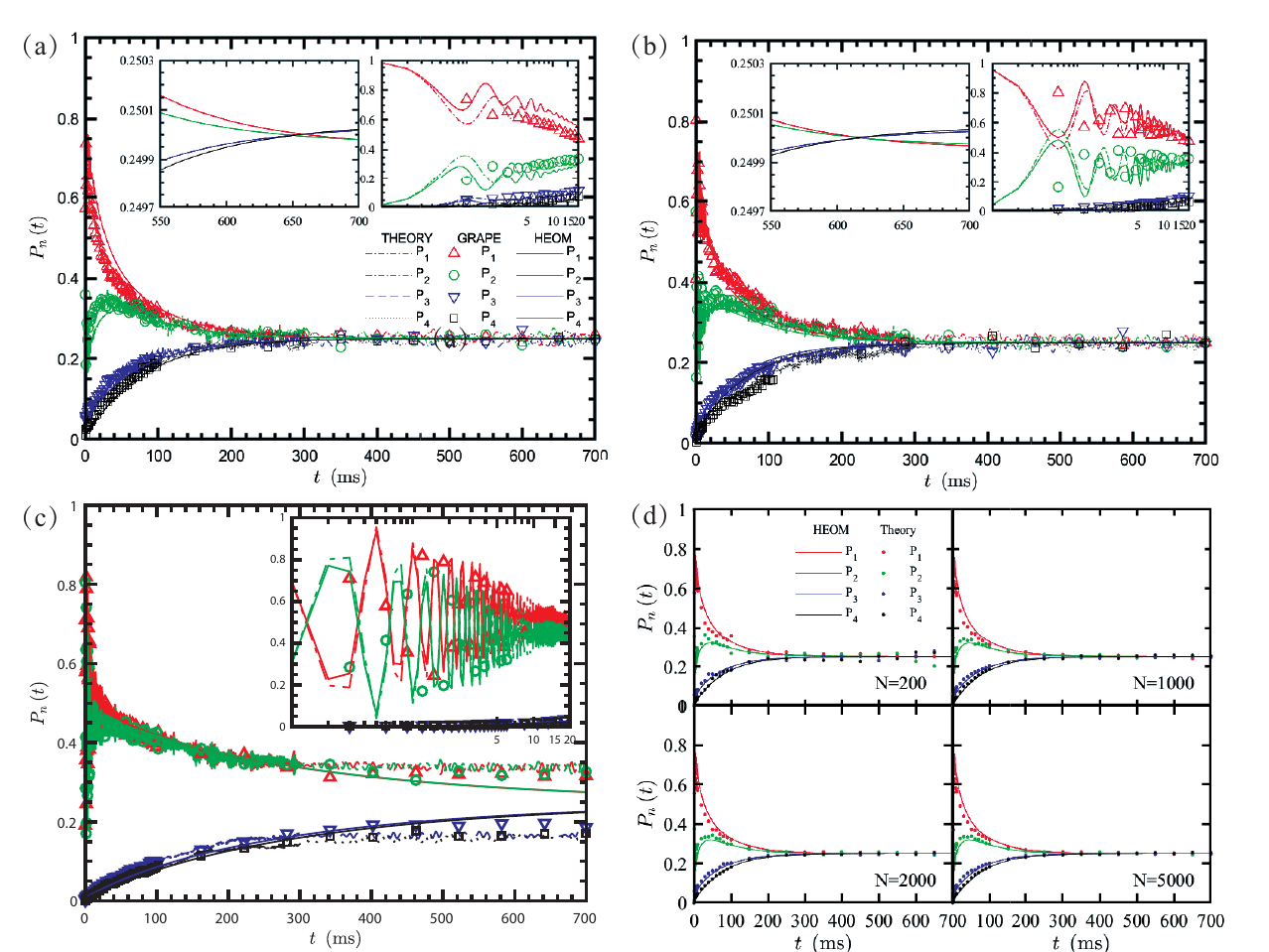}
\caption{Simulations of the energy transfer by the HEOM (curves) and GRAPE (symbols) and quantum simulation (broken curves) with: (a) $r=13.4~\rm{\mathring A}$; (b) $r=11.3~\rm{\mathring A}$; (c) $r=8.0~\rm{\mathring A}$. The last two are averaged over 500 random realizations. The insets enlarge the simulations by the HEOM to show the EET times and the coherent oscillation in the short-time regime. (d) shows the convergence between the HEOM (curves) and quantum simulations (dots) with $r=13.4~\rm{\mathring A}$ as the ensemble size $N$ increases. In all simulations, we take $\gamma_\textrm{NMR}=2\pi\times50~\rm{Hz}$ and $\lambda_\textrm{NMR}=2\pi\times2~\rm{Hz}$. }
\label{fig2}
\end{figure*}

In Fig.~\ref{fig2}(a), we show the comparison among the HEOM, the quantum simulation and the GRAPE simulation for $r=13.4~\rm{\mathring A}$. Therein, all four chlorophylls are equally spaced and the Hamiltonian for the NMR quantum simulation $H_{\rm{NMR}}=H_{\rm{EET}}/C$ is
\begin{equation}
\frac{H_{\rm{NMR}}}{2\pi}={
\left[ \begin{array}{cccc}
130 & 1.2608 & 0.1612 & 0.0474\\
1.2608 & 129 & 1.3190 & 0.1612\\
0.1612 & 1.3190 & 123 & 1.2608\\
0.0474 & 0.1612 & 1.2608 & 122
\end{array}
\right]}~\rm{kHz}.
\end{equation}
In Fig.~\ref{fig2}(b), we set $r=11.3~\rm{\mathring A}$ and thus the distance between the two donors (acceptors) is slightly smaller than the distance between the central two sites. In this case, the donors (acceptors) form a dimer. The corresponding Hamiltonian reads
\begin{equation}
\frac{H_{\rm{NMR}}}{2\pi}={
\left[ \begin{array}{cccc}
130 & 2.1025 & 0.1283 & 0.0474\\
2.1025 & 129 & 0.5759 & 0.1283\\
0.1283 & 0.5759 & 123 & 2.1025\\
0.0474 & 0.1283 & 2.1025 & 122
\end{array}
\right]}~\rm{kHz}.
\end{equation}
In Fig.~\ref{fig2}(c), we further reduce the intra-dimer distance to an even smaller value, i.e., $r=8~\rm{\mathring A}$. Thus, the Hamiltonian is
\begin{equation}
\frac{H_{\rm{NMR}}}{2\pi}={
\left[ \begin{array}{cccc}
130 & 5.9251 & 0.0926 & 0.0474\\
5.9251 & 129 & 0.2195 & 0.0926\\
0.0926 & 0.2195 & 123 & 5.9251\\
0.0474 & 0.0926 & 5.9251 & 122
\end{array}
\right]}~\rm{kHz}.
\end{equation}
In order to obtain the above Hamiltonians, we adopt the site energies $\varepsilon_{1}\!=\!13000~\rm{cm}^{-1}$, $\varepsilon_{2}\!=\!12900~\rm{cm}^{-1}$, $\varepsilon_{3}\!=\!12300~\rm{cm}^{-1}$, and $\varepsilon_{4}\!=\!12200~\rm{cm}^{-1}$, which are typical in natural photosynthetic systems, for the corresponding Hamiltonians of photosynthesis.
Except that $J_{14}=J_{41}$ are invariant, different distances $r$'s correspond to a set of different coupling terms $J_{ij}$'s ($i,j=1,2,3,4$) as determined by Eq.~(\ref{eq:J}).

In addition, we assume the spectral density of Drude-Lorentzian form as given in Eq.~(\ref{eq:Geet}), with the optimal frequency $\gamma_{j}$.
In the simulations, we assume identical phonon relaxation rates $\gamma_\textrm{NMR}=\gamma_\textrm{EET}/C=2\pi\times50~\rm{Hz}$ and identical reorganization energies $\lambda_\textrm{NMR}=\lambda_\textrm{EET}/C=2\pi\times2~\rm{Hz}$ for all local baths. Since the results of quantum simulation coincide with those of the HEOM only at high temperatures, we take $T_{\rm{EET}}=3\times10^{6}$~K and $T_{\rm{NMR}}=10^{-3}~\rm{K}$ in our numerical calculations. In this way, we compare the results of the quantum simulation, and the GRAPE simulation, and the HEOM in Fig.~\ref{fig2}.

\begin{figure}[ht]
\centering
\includegraphics[bb=0 0 270 600,width=6.5cm]{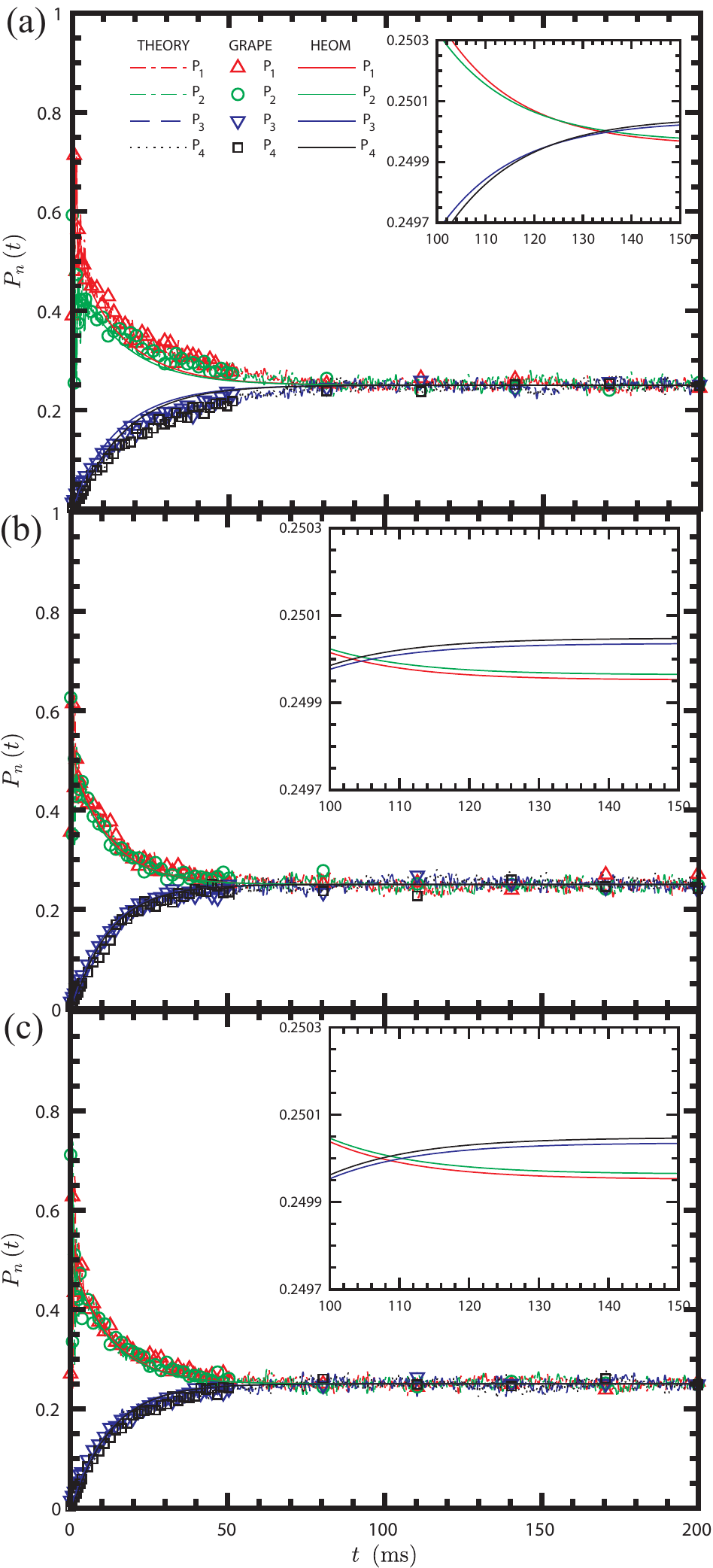}
\caption{Simulations of the energy transfer by the HEOM (curves) and GRAPE algorithm (symbols) with $r=11.3~\rm{\mathring A}$, $\lambda_\textrm{NMR}=2\pi\times2~\rm{Hz}$, and: (a) $\gamma_\textrm{NMR}=2\pi\times0.5~\rm{kHz}$, (b) $\gamma_\textrm{NMR}=2\pi\times2.668~\rm{kHz}$, (c) $\gamma_\textrm{NMR}=2\pi\times7~\rm{kHz}$. The insets show the EET times. }
\label{fig3}
\end{figure}

As shown in Fig.~\ref{fig2}(b), the energy transfer is the fastest for the case with $r=11.3~\rm{\mathring A}$. The dimerized geometry explores the coherent relaxation within the donors to accelerate the energy transfer. However, over-dimerization in the geometry significantly reduces the energy transfer rate, because it enlarges the energy gap between the donor and acceptor clusters, cf. Fig.~\ref{fig2}(c). These numerically-exact simulations are consistent with those discoveries obtained by the approximate theory in Ref.~\cite{Ai13}. We notice that there are some small differences between the HEOM and the quantum simulations, which rely on the assumption that the ensemble average is equivalent to the time average. Therefore, we verify the assumption in Fig.~\ref{fig2}(d). As the number of random realizations in the ensemble increases, the results of quantum simulation approach closer and closer to those of the HEOM. Besides, we also observe the deviations of the GRAPE simulations from the quantum simulations, of which causes will be discussed at the end of this section. In this regard, both the quantum simulations and GRAPE simulations successfully reproduce the coherent oscillations at the short-time regime and the incoherent relaxation at the long-time regime, and it is valid to exactly simulate the open quantum dynamics with a generic Hamiltonian by the above approach.

 In Ref.~\cite{Rey13}, it was shown that the optimization of the EET can be achieved when the energy gap of the system matches the optimum in the spectral density. Inspired by this discovery, we subsequently consider the effects of the bath on the EET dynamics for a fixed Hamiltonian with $r=11.3~\rm{\mathring A}$, which was shown to be optimal among different geometries in Fig.~\ref{fig2}. Since the peak of the Drude-Lorentz spectral density is located at $\gamma_\textrm{NMR}$, we fix $\lambda_\textrm{NMR}$ and simulate the EET dynamics for a broad range of $\gamma_\textrm{NMR}$. Here, in Fig.~\ref{fig3}, we only demonstrate the population dynamics for three typical parameters, i.e., $\gamma_\textrm{NMR}/2\pi=0.5~\rm{kHz}$, $2.668~\rm{kHz}$, and $7~\rm{kHz}$.
Obviously, the optimal relaxation rate of the bath is $\gamma_\textrm{NMR}^\textrm{opt}=2\pi\times2.668~\rm{kHz}$ as it requires the shortest time to achieve equal populations.
The physical mechanism can be described by the energy diagram of the system.
Due to their strong couplings, sites 1 and 2 form the donor cluster, while sites 3 and 4 form the acceptor cluster. Since the donor cluster is weakly coupled with the acceptor cluster, the EET between the two clusters can be described by the F\"{o}rster theory.
Therefore, the inter-cluster EET is optimized when the energy gap between the lower-energy eigen-state of the donor cluster and the higher-energy eigen-state of the acceptor cluster matches the optimum of the spectral density.
The details about this physical mechanism is elucidated in Appendix~\ref{sec:AppendixB}.

\subsection{Different Types of Spectral Densities}

Since a general power-law form spectral density can describe an extremely-large number of physical environments \cite{Weiss08,Leggett87}, we use it to show the universal applicability of the quantum simulation approach. Following Refs.~\cite{Breuer07,Chin12}, we analyze the EET dynamics for three types of spectral densities, namely Ohmic, sub-Ohmic, and super-Ohmic spectral densities, which are expressed in a unified manner as
\begin{equation}
J(\omega)=\frac{\lambda\omega}{\Gamma(s)}\left(\frac{\omega}{\omega_{c}}\right)^{s-1}e^{-\frac{\omega}{\omega_{c}}},
\label{eq:JOhmic}\end{equation}
where $\Gamma(s)$ is the Euler Gamma function, and $\omega_{c}$ is an exponential cutoff frequency, $\lambda\omega_{c}$ is the coupling strength between the system and the bath. When $0<s<1$, $s=1$, and $s>1$, $J(\omega)$ denotes sub-Ohmic, Ohmic, and super-Ohmic spectral densities, respectively.

Here, we take $s=1,0.5,3$ corresponding to respectively the Ohmic, sub-Ohmic, and super-Ohmic spectral density, i.e.,
\begin{subequations}
\begin{align}
&J_\textrm{Ohm}(\omega)=\lambda\omega e^{-\frac{\omega}{\omega_{c}}},\\
&J_\textrm{sub}(\omega)=\frac{\lambda\omega}{\pi^{1/2}}\left(\frac{\omega}{\omega_{c}}\right)^{-1/2} e^{-\frac{\omega}{\omega_{c}}},\\
&J_\textrm{sup}(\omega)=\frac{\lambda\omega^{3}}{2\omega_{c}^{2}} e^{-\frac{\omega}{\omega_{c}}}.
\end{align}
\end{subequations}
Through the analyses in the above subsection, we can see that the results of the quantum simulations are very coincident with those of the HEOM at high temperatures. In the following, we also analyze the EET dynamics for these three types of spectral densities.


\begin{figure*}[htbp]
\centering
\includegraphics[scale=0.94]{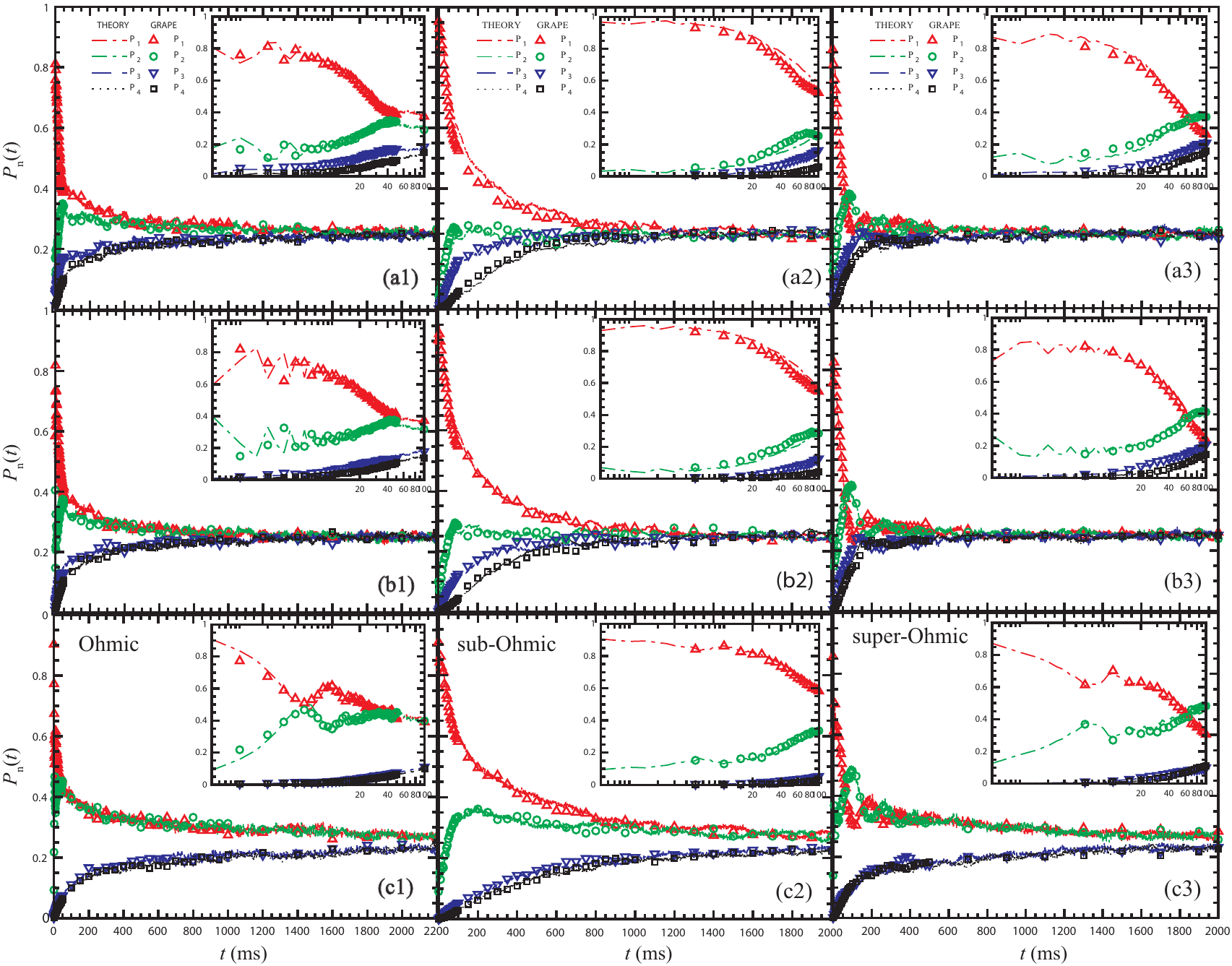}
\caption{Simulations of the energy transfer for the Ohmic, sub-Ohmic, and super-Ohmic spectral densities by the quantum simulation (curves) and GRAPE algorithm (symbols) with $N=500$. (a1), (b1), (c1) represent the results of Ohmic spectral density, with $\lambda=10^{-3}$, $\omega_{c}=2\pi\times100~\rm{Hz}$, and $\omega_{0}=2\pi\times0.1~\rm{Hz}$. (a2), (b2), (c2) show the results of sub-Ohmic spectral density, with $\lambda=0.05$, $\omega_{c}=2\pi\times50~\rm{Hz}$, and $\omega_{0}=2\pi\times0.1~\rm{Hz}$. (a3), (b3), (c3) reveal the results of super-Ohmic spectral density, with $\lambda=0.05$, $\omega_{c}=2\pi\times40~\rm{Hz}$, and $\omega_{0}=2\pi\times0.1~\rm{Hz}$. The three rows correspond to $r=13.4~\rm{\mathring A}$, $r=11.3~\rm{\mathring A}$, and $r=8.0~\rm{\mathring A}$, respectively.}
\label{fig44}
\end{figure*}

Figure~\ref{fig44} presents the results of our quantum simulations for Ohmic, sub-Ohmic and super-Ohmic spectral densities, with different geometries, i.e., $r=13.4~\rm{\mathring A}$, $r=11.3~\rm{\mathring A}$, and $r=8~\rm{\mathring A}$. In the numerical calculations, we take the temperature as $T_\textrm{EET}=3\times10^{5}~\rm{K}$.
Interestingly, we can see that the EET dynamics is strongly dependent on the type of spectral density adopted, i.e., the statistics of the system-bath interactions. Therefore, it is crucial a simulation method can accurately simulate all types of spectral densities. Nevertheless, regardless of the form of the spectral densities, the system identically reaches the equilibrium fastest when the intra-pair distance is $r=11.3~\rm{\mathring A}$, cf. Figs.~\ref{fig44}(b1),(b2),(b3), compared with the other distances. As $r$ is reduced, the coherent oscillations in the populations of the donors becomes more and more profound, but the EET times do not decrease monotonically.
For these three spectral densities, we may arrive at the same conclusion as the Drude-Lorentzian spectral density, that the moderate-dimerized geometry explores the coherent relaxation within the donors to accelerate the energy transfer, and  the over-dimerization in the geometry significantly reduces the energy transfer rate, cf. Figs.~\ref{fig44}(c1),(c2),(c3), which is also consistent with the conclusion in Ref.~\cite{Ai13}.

Based on the above analysis, we demonstrate that this quantum simulation approach can be used to investigate the exact quantum dynamics for different Hamiltonians and various types of spectral densities. It has been proven that although theoretically the quantum dynamics with arbitrary form of spectral density can be simulated by the HEOM through spectral decomposition \cite{Liu14,Schroder07}, this may be practically unfeasible. As will be shown in the next subsection, the more complex the form of spectral density is, the higher the computational complexity of the HEOM will be. However, since the computational time of the present simulation approach is not affected by the complexity of the spectral density, we show the superiority of our method over the HEOM in the high-temperature limit.

\color{black}
\subsection{Computational Costs of NMR Simulation and HEOM}

In the above sections, we show the quantum simulations of open quantum dynamics for a 4-level system. Consider an $N$-level open quantum system, where each level of the system is coupled with an independent bath, and the correlation function of each bath contains $K$ exponentials. The total number of auxiliary density operators for the HEOM is \cite{Shi09,Ishizaki09-3}
\begin{equation}
\mathcal{I}=\sum_{n=0}^{\mathcal{N}}\mathcal{I}_n=\sum_{n=0}^{\mathcal{N}}\frac{(n+KN-1)!}{n!(KN-1)!}=\frac{(\mathcal{N}+KN)!}{\mathcal{N}!(KN)!},\label{eq:I}
\end{equation}
where $\mathcal{N}$ is the hierarchy level of truncation of the HEOM.
By using \cite{Robbins1955}
\begin{equation}
n!=\sqrt{2\pi}n^{n+\frac{1}{2}}e^{-n}e^{r_n},
\end{equation}
where $(12n+1)^{-1}<r_n<(12n)^{-1}$, we can obtain Eq.~(\ref{eq:I}) as
\begin{eqnarray}
\mathcal{I}_n&=&\sqrt{\frac{\frac{1}{\mathcal{N}}+\frac{1}{KN}}{2\pi}} \left(1+\frac{KN}{\mathcal{N}}\right)^{\mathcal{N}}\!\!
\left(1+\frac{\mathcal{N}}{KN}\right)^{KN}\nonumber \\
&&\times\exp(r_{\mathcal{N}+KN}-r_{\mathcal{N}}-r_{KN}).
\end{eqnarray}
When the dimension of the system is large and the spectral density is complex, i.e., $KN\rightarrow\infty$, for a given hierarchy level of truncation, the total number of auxiliary density operators and thus the computation cost of the HEOM is approximated as
\begin{eqnarray}
\lim \limits_{KN\to \infty}\mathcal{I}\simeq\sqrt{\frac{1}{2\pi\mathcal{N}}}\left(1+\frac{KN}{\mathcal{N}}\right)^{\mathcal{N}}e^{KN}.
\end{eqnarray}

On the other hand, for a $\log_2{N}$-qubit quantum system, the GRAPE algorithm requires \cite{Li17}
\begin{equation}
(4M\log_2{N}+1)4^{\log_2{N}}=(4M\log_2{N}+1)N^{2}
\end{equation}
measurements in each iteration to estimate the fitness function and its derivative with respect to the pulse amplitude, which the GRAPE aims to optimize. $M$ is the number of the pulse sequence to be divided, which scales polynomially with the number of qubits \cite{Li17}. Above all, the complexity of the GRAPE algorithm is a polynomial of $N$.

In practical simulations, the resources required by the HEOM may be intolerable as complicated spectral densities are widely observed in natural photosynthetic complexes. According to Ref.~\cite{Novoderezhkin04}, there are 42 chlorophylls in a trimer of LHC II complex and its spectral density can be described by an overdamped Brownian oscillator and 48 high-frequency modes, i.e., $N=42$ and $K=49$.
We remark that because of the specific mapping from $N$-level photosynthetic energy transfer to $\log_2{N}$-qubit quantum simulation, the complexity has been greatly reduced.

\subsection{Error Analysis}

From the above theoretical calculations, we can see that there are some errors in the quantum simulations as compared to the theoretically-exact results by the HEOM, especially when the number of random realizations in the ensemble is small. Theoretically, the average over the ensemble is equal to the average over the time only in the infinite-large ensemble limit \cite{Goodman15}. However, according to a large number of simulations, the quantum simulations agree well with the HEOM for $n\geq500$. For a given random Hamiltonian, the corresponding unitary evolution can be decomposed into a sequence of experimentally-feasible pulses by the GRAPE algorithm. Here, the fidelity is limited by the initial guess of parameters, in combination with both the step and the number of repetitions in attaining the global optimum \cite{Li17}.

In the aspect of NMR realization, there are three main sources of errors. First of all, the prepared initial state is a pseudo-pure state rather than a pure state. Then, in the process of quantum-state evolution, although the fidelity of the unitary evolution $U_D$ calculated by the GRAPE algorithm can in principle approach unity, e.g. by increasing the step of the repetition, there could still remain imperfection in the experimental realization of the pulse sequence. Finally, further errors could be introduced in the process of measurement.


\section{Conclusion}
\label{sec:Conclusion}

In this paper, we discuss the recently-developed approach for the exact simulation of EET dynamics in photosynthesis. By applying the approach to the linear-tetramer model, the energy transfer is shown to be optimal for a moderately-clustered geometry. Based on the optimal Hamiltonian, we show that the energy transfer efficiency can be further improved when the energy gap between the donor and acceptor clusters matches the optimum of the bath's spectral density. In this regard, we demonstrate that the light-harvesting network can be optimized from two aspects, i.e., the geometry \cite{Ai13} and the bath \cite{Rey13}.

Beyond the Drude-Lorentz spectral density, we also show that our approach can be utilized to simulate the EET dynamics for various types of spectral densities. By comparing our approach to the HEOM, the complexity is a polynomial of the number of states involved and thus our approach is exponentially accelerated when the system is large and the spectral density is complicated. This is often encountered when simulating open quantum dynamics for natural photosynthesis. To conclude, our approach can be applied to the exact and efficient simulation of open quantum dynamics for various of Hamiltonians and spectral densities. It may shed light on the investigations exploring non-Markovianity in open quantum dynamics, e.g. quantum metrology in non-Markovian environments \cite{Chin12}.

\begin{acknowledgments}
We thank the critical comments from J.-S. Shao, and valuable discussions with B. X. Wang and J. W. Wen.
This work is supported by the National Natural Science Foundation of China under Grant Nos.~11674033,~11474026,~11505007, and Beijing Natural Science Foundation under Grant No.~1202017.
N.L. acknowledges partial support from JST PRESTO through Grant No.~JPMJPR18GC.
\end{acknowledgments}

\appendix
\section{Measuring Off-diagonal Terms of Density Matrix}
\label{sec:AppendixA}

Concerning 2 qubits, there are 9 independent elements of the entire density matrix, including 3 diagonal terms and 6 off-diagonal terms. In Sec.~\ref{subsec:Tomography}, we introduce how to measure the diagonal elements of the density matrix, i.e., the populations. In order to perform the tomography, we will supplement the method for measuring the off-diagonal elements, which can be given as follows, i.e.,
\begin{align}
\rho_{21}&\!\!=\!\!\frac{1}{4}[\langle\sigma^{(0)}_1\sigma^{(1)}_2\rangle+\langle \sigma^{(3)}_1\sigma^{(1)}_2\rangle+i(\langle \sigma^{(0)}_1\sigma^{(2)}_2\rangle+\langle \sigma^{(3)}_1\sigma^{(2)}_2\rangle)],\notag\\
\rho_{31}&\!\!=\!\!\frac{1}{4}[\langle\sigma^{(1)}_1\sigma^{(0)}_2\rangle+\langle \sigma^{(1)}_1\sigma^{(3)}_2\rangle+i(\langle \sigma^{(2)}_1\sigma^{(0)}_2\rangle+\langle \sigma^{(2)}_1\sigma^{(3)}_2\rangle)],\notag\\
\rho_{41}&\!\!=\!\!\frac{1}{4}[\langle\sigma^{(1)}_1\sigma^{(1)}_2\rangle-\langle \sigma^{(2)}_1\sigma^{(2)}_2\rangle+i(\langle \sigma^{(1)}_1\sigma^{(2)}_2\rangle+\langle \sigma^{(2)}_1\sigma^{(1)}_2\rangle)],\notag\\
\rho_{32}&\!\!=\!\!\frac{1}{4}[\langle\sigma^{(1)}_1\sigma^{(1)}_2\rangle+\langle \sigma^{(2)}_1\sigma^{(2)}_2\rangle-i(\langle \sigma^{(1)}_1\sigma^{(2)}_2\rangle-\langle \sigma^{(2)}_1\sigma^{(1)}_2\rangle)],\notag\\
\rho_{42}&\!\!=\!\!\frac{1}{4}[\langle\sigma^{(1)}_1\sigma^{(0)}_2\rangle-\langle \sigma^{(1)}_1\sigma^{(3)}_2\rangle+i(\langle \sigma^{(2)}_1\sigma^{(0)}_2\rangle-\langle \sigma^{(2)}_1\sigma^{(3)}_2\rangle)],\notag\\
\rho_{43}&\!\!=\!\!\frac{1}{4}[\langle\sigma^{(0)}_1\sigma^{(1)}_2\rangle-\langle \sigma^{(3)}_1\sigma^{(1)}_2\rangle+i(\langle \sigma^{(0)}_1\sigma^{(2)}_2\rangle+\langle \sigma^{(3)}_1\sigma^{(2)}_2\rangle)].
\label{eq:A1}
\end{align}
In order to obtain all the elements of the density matrix, we apply a unitary operation $U$ to the system with $U=\{
\sigma_1^{(0)} \otimes \sigma_2^{(0)},
\sigma_1^{(0)} \otimes e^{-i\frac{\pi}{4}\sigma_2^{(2)}},
e^{-i\frac{\pi}{4}\sigma_1^{(2)}} \otimes \sigma_2^{(0)},
e^{-i\frac{\pi}{4}\sigma_1^{(2)}} \otimes e^{-i\frac{\pi}{4}\sigma_2^{(2)}}
\}$.
In addition to Eq.~(\ref{eq:diagObser}), the expectations of the other 12 observables are respectively written in terms of the FID signals as
\begin{subequations}
\begin{align}
\langle\sigma^{(0)}_1\sigma^{(1)}_2\rangle=\frac{\eta}{2}[\textrm{Re}(S_{20}^{\textrm{YI}})+\textrm{Re}(S_{21}^{\textrm{YI}})],\\
\langle\sigma^{(0)}_1\sigma^{(2)}_2\rangle=\frac{\eta}{2}[\textrm{Im}(S_{20}^{\textrm{YI}})+\textrm{Im}(S_{21}^{\textrm{YI}})],\\
\langle\sigma^{(1)}_1\sigma^{(0)}_2\rangle=\frac{\eta}{2}[\textrm{Re}(S_{10}^{\textrm{IY}})+\textrm{Re}(S_{11}^{\textrm{IY}})],\\
\langle\sigma^{(1)}_1\sigma^{(1)}_2\rangle=\frac{\eta}{2}[\textrm{Re}(S_{11}^{\textrm{IY}})-\textrm{Re}(S_{10}^{\textrm{IY}})],\\
\langle\sigma^{(1)}_1\sigma^{(2)}_2\rangle=\frac{\eta}{2}[\textrm{Im}(S_{21}^{\textrm{YI}})-\textrm{Im}(S_{20}^{\textrm{YI}})],\\
\langle\sigma^{(1)}_1\sigma^{(3)}_2\rangle=\frac{\eta}{2}[\textrm{Re}(S_{10}^{\textrm{XI}})-\textrm{Re}(S_{11}^{\textrm{XI}})],\\
\langle\sigma^{(2)}_1\sigma^{(0)}_2\rangle=\frac{\eta}{2}[\textrm{Im}(S_{10}^{\textrm{IY}})+\textrm{Im}(S_{11}^{\textrm{IY}})],\\
\langle\sigma^{(2)}_1\sigma^{(1)}_2\rangle=\frac{\eta}{2}[\textrm{Re}(S_{11}^{\textrm{IY}})-\textrm{Re}(S_{10}^{\textrm{IY}})],\\
\langle\sigma^{(2)}_1\sigma^{(2)}_2\rangle=\frac{\eta}{2}[\textrm{Im}(S_{20}^{\textrm{XI}})-\textrm{Im}(S_{21}^{\textrm{XI}})],\\
\langle\sigma^{(2)}_1\sigma^{(3)}_2\rangle=\frac{\eta}{2}[\textrm{Im}(S_{10}^{\textrm{YI}})-\textrm{Im}(S_{11}^{\textrm{YI}})],\\
\langle\sigma^{(3)}_1\sigma^{(1)}_2\rangle=\frac{\eta}{2}[\textrm{Re}(S_{20}^{\textrm{IX}})-\textrm{Re}(S_{21}^{\textrm{IX}})],\\
\langle\sigma^{(3)}_1\sigma^{(2)}_2\rangle=\frac{\eta}{2}[\textrm{Im}(S_{20}^{\textrm{IY}})-\textrm{Im}(S_{21}^{\textrm{IY}})].
\end{align}
\end{subequations}

\section{Optimizing EET by Bath}
\label{sec:AppendixB}

In Fig.~\ref{fig3}, we show that the optimization of the EET can be also achieved when the energy gap of the system matches the optimum in the spectral density.
In order to elucidate the underlying physical mechanism, we resort to the energy diagram of the system.
Sites 1 and 2 form the donor cluster due to their strong coupling, while sites 3 and 4 form the acceptor cluster as shown in Fig.~\ref{fig1}(a).
By diagonalizing the Hamiltonians of donor and acceptor subsystems respectively, i.e.,
\begin{subequations}
\begin{align}
H_{\rm{S}}^{12}&=\varepsilon_{1}\vert 1\rangle\langle 1\vert+\varepsilon_{2}\vert 2\rangle\langle 2\vert+J_{12}\vert 1\rangle\langle 2\vert+J_{21}\vert 2\rangle\langle 1\vert, \\
H_{\rm{S}}^{34}&=\varepsilon_{3}\vert 3\rangle\langle 3\vert+\varepsilon_{4}\vert 4\rangle\langle 4\vert+J_{34}\vert 3\rangle\langle 4\vert+J_{43}\vert 4\rangle\langle 3\vert,
\end{align}
\end{subequations}
we can obtain the eigen-states as
\begin{subequations}
\begin{align}
\vert E_{1}\rangle&=\cos\frac{\theta_{12}}{2}\vert 1\rangle+\sin\frac{\theta_{12}}{2}\vert 2\rangle, \\
\vert E_{2}\rangle&=\sin\frac{\theta_{12}}{2}\vert 1\rangle-\cos\frac{\theta_{12}}{2}\vert 2\rangle, \\
\vert E_{3}\rangle&=\cos\frac{\theta_{34}}{2}\vert 3\rangle+\sin\frac{\theta_{34}}{2}\vert 4\rangle, \\
\vert E_{4}\rangle&=\sin\frac{\theta_{34}}{2}\vert 3\rangle-\cos\frac{\theta_{34}}{2}\vert 4\rangle,
\end{align}
\end{subequations}
where $\theta_{j,j+1}=\arctan\left(\frac{2J_{j,j+1}}{\varepsilon_{j}-\varepsilon_{j+1}}\right)~(j=1,3)$ are the mixing angles.
And the corresponding eigen-energies are respectively
\begin{subequations}
\begin{align}
E_{1}&=\frac{\varepsilon_{1}+\varepsilon_{2}}{2}+\sqrt{\left(\frac{\varepsilon_{1}-\varepsilon_{2}}{2}\right)^{2}+J_{12}^{2}},\\
E_{2}&=\frac{\varepsilon_{1}+\varepsilon_{2}}{2}-\sqrt{\left(\frac{\varepsilon_{1}-\varepsilon_{2}}{2}\right)^{2}+J_{12}^{2}},\\
E_{3}&=\frac{\varepsilon_{3}+\varepsilon_{4}}{2}+\sqrt{\left(\frac{\varepsilon_{3}-\varepsilon_{4}}{2}\right)^{2}+J_{34}^{2}},\\
E_{4}&=\frac{\varepsilon_{3}+\varepsilon_{4}}{2}-\sqrt{\left(\frac{\varepsilon_{3}-\varepsilon_{4}}{2}\right)^{2}+J_{34}^{2}}.
\end{align}
\end{subequations}
Because the donor and acceptor clusters are weakly coupled, the F\"{o}rster mechanism can be utilized to describe the inter-cluster EET. As a result,
the optimal $\gamma^\textrm{opt}_\textrm{NMR}$ coincides with the energy gap between the lower eigen-state of the donor cluster and the higher eigen-state of the acceptor cluster, i.e., $\gamma^\textrm{opt}_\textrm{NMR}=E_2-E_3$. Above all, based on the optimal geometry, we can further optimize the energy-transfer efficiency by tuning the bath to match the energy gap of the system.


\end{document}